\documentclass[english,aps,manuscript]{revtex4}
\usepackage[T1]{fontenc}
\usepackage[latin9]{inputenc}
\usepackage{graphicx}
\usepackage{amssymb}

\makeatletter

\providecommand{\boldsymbol}[1]{\mbox{\boldmath $#1$}}

\providecommand{\tabularnewline}{\\}



\usepackage{babel}
\makeatother

\begin{document}

\title{Classification of the Super-conducting order parameters under the
point group symmetry for a multi-band system: application to LaOFeAs }

\author{Zhi-Hui Wang$^{1}$, Hui Tang$^{2,1}$, Zhong Fang$^{2}$ and Xi
Dai$^{2}$}

\affiliation{$^{1}$Institute of Theoretical Physics, Chinese Academy of Sciences,
Beijing 100190, China}

\affiliation{$^{2}$Beijing National Laboratory for Condensed Matter Physics,
and Institute of Physics, Chinese Academy of Sciences, Beijing 100190,
China}

\begin{abstract}
All the possible super-conducting order parameters for the LaOFeAs
system are classified by their transformation under the complete crystal
symmetry. The general forms of the super-conducting gap functions
for each class are discussed. We find that the gap functions in such
a multi-band system belong to three types, full gap, nodal type and
finite {}``Fermi arc'' type. Possible physical consequences caused
by different types of gap functions are also discussed. 
\end{abstract}

\email{daix@aphy.iphy.ac.cn}

\maketitle

\section{Introduction}

The iron based superconductors, the second family of the high temperature
superconducting materials, have been discovered very recently. Hosono
group first obtained super-conductivity in LaOFeAs with $T_{c}=26K$
by replacing some Oxygen atom with F~\citep{LOFS}. Soon after it,
super-conductivity is discovered in several similar materials with
$T_{c}=41$ K in $CeO_{1-x}F_{x}FeAs$~\citep{COFS} and $T_{c}=43$
K in $SmO_{1-x}F_{x}FeAs$~\citep{SOFS}. Further replacing $La$
atoms with other rare earth elements rapidly raises $T_{c}$ up to
$52K$ in $Pr[O_{1-x}F_{x}]FeAs$~\citep{POFS} and $55K$ in $Sm[O_{1-x}F_{x}]FeAs$~\citep{SOFS2}.
With some other ways to introduce carriers, super-conductivity with
the $T_{c}$$=55K$ in $Gd_{1-x}Th_{x}OFeAs$ and $ReFeAsO_{1-\delta}$
are obtained by several groups~\citep{GOFS,SOxFS2,Hole}.

N. L. Wang's and Z. Fang's group first reported that the ground state
of the parent compound shows SDW long range order with stripe like
spin configuration by combining the first principle calculation and
optical measurement~\citep{SDW}. The SDW state has been confirmed
by neutron scattering results from two independent groups~\citep{Neutron,Neutron2},
and several other calculations\citep{SDW-Cal,SDW-Cal2}. The NMR data
also show strong evidence of magnetic phase transition at 135K~\citep{NMR,NMR2},
which is consistent with the previous neutron scattering and optical
conductivity experiments. Upon $F$ doping, the SDW is suppressed
very quickly and superconductivity appears. Since the maximum $T_{c}$
achieved in this family is around 55K, which is well above the McMillan
limit, the super-conductivity here is likely to be non-BCS type.

The pairing symmetry of the super-conductivity in iron based super-conductors
is one of the key issues. The strong magnetic fluctuation, which is
suggested by the first principle calculation~\citep{SDW,Xu,Singh,Mazin,LOFP-cal,Kotliar,Kuroki}
and NMR measurement~\citep{NMR,NMR2}, is considered by many authors
to be the {}``pairing glue''. However it is still under debate if
the ferromagnetic\citep{Singh,Xu,HJZhang} or anti-ferromagnetic\citep{SDW-Cal,SDW-Cal2}
fluctuation is responsible for the super-conductivity here. The ferromagnetic
(anti-ferromagnetic) fluctuation will mediate attractive interaction
in the spin triplet (singlet) channel and thus leads to the triplet
(singlet) super-conducting state\citep{Dai,Qimiao,Scalapino,Shoucheng,Zidan,Xiaogang}.
The current experimental data still can not confirm the spin state
of the Cooper pairs and band structure calculation indicates nearly
degenerate multiple Fermi Surfaces (FS) in the system~\citep{HJZhang}.
Therefore at the current stage, there still exist many possible pairing
order parameters leading to many different types of gap functions~.
It is then important to classify all these possible super-conducting
order parameters by their different transformation behavior under
the full symmetry of the crystals, which will be done in the following
sections. After the classification, we will obtain many classes of
the super-conducting order parameters with each of them forming an
irreducible representation of the full symmetry the system have. Also
we will discuss the general form of the gap functions for each class
of the super-conducting phase, which can be grouped into three catalogs.
The simplest type, which is referred to type (I) gap function, has
a fully opened gap over the whole FS for any non-zero order parameter.
The type (II) gap function contains a nodal point on the FS where
it vanishes. And the type (III) gap function will vanish on a finite
section of the FS, which is mentioned as {}``Fermi arc'' in the
previous literature~\citep{Dai}.

\section{Classification of the super-conducting order parameters }

The full symmetry of the crystal can be view as the direct production
of time reversal, spacial inversion, charge conservation,translational
symmetry and the point group symmetry. In the present study, we assume
that the appearance of super-conductivity only breaks the charge conservation
and keeps all the other symmetry. The purpose of the current paper
is to classify all the possible super-conducting order parameters
by the group representation theory.

\subsection{The Fermi surface topology and the point group symmetry of $LaOFeAs$}

In the absence of SDW state (after F-doping), LaOFeAs crystallizes
in layered square lattice with space group P4/nmm. There are two Fe
sites per unit cell, which are coordinated by As-tetrahedrons. The
point group symmetry of $LaOFeAs$ is $D_{4h}$, which contains a
4-fold rotational axis along z-direction, four vertical reflection
planes along $x,y,xy$ and $\bar{x}y$ directions, a horizontal reflection
plane, as well as the spacial inversion. According to the first principle
calculation, for non magnetic $LaOFeAs$, \citep{Xu,Singh,Mazin,LOFP-cal,Kotliar,Kuroki}there
are two hole type Fermi surfaces enclosing $\Gamma$ point and two
electron type Fermi surfaces enclosing $M$ point. Upon $F$ doping,
the three hole type Fermi surfaces shrink rapidly and will be neglected
in the present paper. Therefore we can focus on the super-conducting
order parameters on the two electron type Fermi surfaces. The shape
of the Fermi surfaces enclosing $M$ point is very close to ellipse
with their long axis along $\left(1,1\right)$ or $\left(\bar{1},1\right)$
directions. The first principle calculation indicates that the states
around the FS around $M$ point are mainly from the $xy,xz$and $yz$
iron 3d orbitals with very strong angle dependence along the FS~\citep{HJZhang}.
Before we begin to classify all the possible superconducting order
parameter $\Delta\left(\boldsymbol{\mathbf{k}}\right)$for a multi-band
system, we first have to define the Bloch states $\psi_{n\mathbf{\boldsymbol{k}}}\left(r\right)$
around $M$ point properly to guarantee$\psi_{n\mathbf{\boldsymbol{k}}}\left(r\right)$
to be smoothly connected along the momentum space. This {}``smooth
condition'' give us unique definition of the energy bands, which
gives the ellipse FS for band {}``1'' ( {}``2'') along $\left(1,1\right)$
($\left(\bar{1},1\right)$ )directions as shown in Fig. 1.

By definition the origin for all the point group operations and spacial
inversion should be located at $\Gamma$ point. While one can easily
prove that using the periodicity of the momentum space, the origin
can be moved to $M$ point. Therefore in the following of the paper,
all the point group and spacial inversion operations are taken respect
to the $M$ point.

\begin{figure}
\begin{centering}
\includegraphics[scale=0.6]{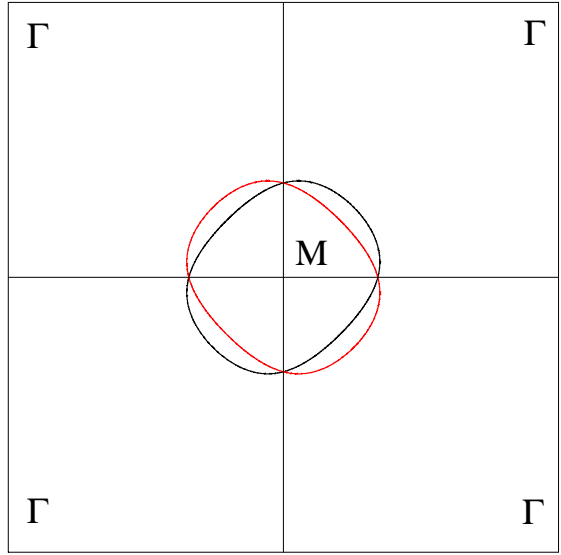}
\par\end{centering}

\caption{The schematic Fermi surface of the LaOFeAs, the FS of band 1 (2) is
plotted by the black (red) ellipse. }

\end{figure}

With such a smooth condition for the energy bands, we can write down
the effect of the $D_{4h}$ group elements $\hat{g}_{l}$ acting on
the Bloch functions in the two bands forming the FS around the $M$
point. In general it can be written as

\[
\hat{g}_{l}\psi_{n\mathbf{\boldsymbol{k}}}\left(r\right)=\hat{D}_{mn}^{orb}\left(\hat{g}_{l}\right)\psi_{n\mathbf{\hat{D}^{R(-)}\left(\hat{g}_{l}\right)\boldsymbol{k}}}\left(r\right)\]

where $2\times2$ matrices $\hat{D}_{mn}^{orb}\left(\hat{g}_{l}\right)$
are the representation matrices in the orbital space, which equal
to unit matrix for operators $\left(E,C_{2}^{"},i,2\sigma_{v}^{"}\right)$and
$\sigma_{x}$ for the rest of them. The above representation in the
orbital space can be checked by the symmetry right on the FS. From
Fig.1, one can easily find that the FS of band {}``1'' ({}``2'')
remains unchanged under the first catalog of the operators and transforms
to FS of band {}``2'' ({}``1'') under the second catalog of the
operators.

\subsection{The super-conducting order parameter and its transformation under
$D_{4h}$}

In the multi-band system, the super-conducting order parameter can
be written as

\[
\Delta_{\mu\nu}^{\alpha\beta}\left(k\right)=\left\langle C_{\alpha\mu}\left(k\right)C_{\beta\nu}\left(-k\right)\right\rangle \]

where $\alpha,\beta=1,2$ are the band Indies, $\mu,\nu=\downarrow,\uparrow$
denote the electron spin. The functions $\Delta_{\mu\nu}^{\alpha\beta}\left(k\right)$
form a representation of the full symmetry of the system, which can
be expressed as the direct production of time reversal, spacial inversion
and point group symmetry. In addition, the anti-symmetric nature of
the fermionic many body wave function requires $\Delta_{\mu\nu}^{\alpha\beta}\left(k\right)=-\Delta_{\nu\mu}^{\beta\alpha}\left(-k\right)$.
In the present study, we follow the treatment of Sigrist and Ueda
on the effect of spin-orbital coupling (SOC)\citep{Sigrist}, assuming
the SOC is not too strong so that the induced coupling between the
singlet and triplet pairing states can be safely neglected. Therefore
we can first divide $\Delta_{\mu\nu}^{\alpha\beta}\left(k\right)$
into singlet and triplet two subgroups. Among them, the singlet pairing
order parameter can be written as a $2\times2$ matrix $\Delta_{singlet}^{\alpha\beta}\left(k\right)$
with the following requirement from the anti-symmetric condition upon
particle exchange,

\[
\Delta_{singlet}^{\alpha\beta}\left(k\right)=\Delta_{singlet}^{\beta\alpha}\left(-k\right)\]
 . Therefore for even (odd) parity state $\Delta_{singlet}^{\alpha\beta}\left(k\right)$
must be a symmetric (anti-symmetric) matrix. Following the standard
notation the order parameters for the triplet pairing state\citep{Leggett}
can be expressed by a vector $\overrightarrow{d}{}^{\alpha\beta}\left(k\right)$with
each component of the d-vector to be a $2\times2$ matrix. And the
requirement from the anti-symmetric condition here is

\[
\overrightarrow{d}{}^{\alpha\beta}\left(k\right)=-\overrightarrow{d}{}^{\beta\alpha}\left(-k\right)\]
 , where for even (odd) parity states $\overrightarrow{d}{}^{\alpha\beta}\left(k\right)$
must be anti-symmetric (symmetric) matrices. With the above notation
of the super-conducting order parameters for both singlet and triplet
pairing states, we can write down the symmetry transformations of
these order parameters under time reversal and point group operators.
For time reversal symmetry we have

\[
K\Delta_{singlet}^{\alpha\beta}\left(k\right)=\Delta_{singlet}^{\alpha\beta*}\left(-k\right)\]
 , for spin singlet states and

\[
K\overrightarrow{d}{}^{\alpha\beta}\left(k\right)=-\overrightarrow{d}{}^{\alpha\beta*}\left(-k\right)\]
 .

At last we consider the symmetry transformation under point group
operators, which are group elements of $D_{4h}$. As we have mentioned
before, in the present study, the energy scale of spin-orbital coupling
is considered to be small compared with the energy splitting between
the spin spin singlet and triplet pairing states but is still big
enough to lift the degeneracy in the spin space spanned by vector
$\overrightarrow{d}{}^{\alpha\beta}$. As the consequence, the spin
degree of freedom of the Cooper pairs has to be frozen to the spacial
motion of the electrons. Therefore the d-vector has to rotate exactly
following the momentum $\boldsymbol{k}$ under the point group operators,
which leads to

\[
\hat{g}_{l}\overrightarrow{d}{}^{\alpha\beta}\left(\boldsymbol{k}\right)=\hat{D}^{R(+)}\left(\hat{g}_{l}\right)\overrightarrow{d}{}^{\alpha\beta}\left(\hat{D}^{R(-)}\left(\hat{g}_{l}\right)\boldsymbol{k}\right)\]

where $\hat{D}^{R(\pm)}$ is the representation in three-dimensional
space with positive (for spin space) or negative (for k-space) spacial
inversion operation respectively.

By the transformation under symmetry operators, we can construct a
representation of the full crystal symmetry using the super-conducting
order parameters as the basis. As we have mentioned above, the two
corresponding Bloch bands we are interested in are defined smoothly
in k-space. Therefore the SC order parameters $\triangle_{\mu\nu}^{\alpha\beta}\left(k\right)$
should be also smooth function in k-space, which allows us to expand
them by first three sphere harmonics and use them as the basis for
k dependence.

In general such a representation is reducible and can be reduced to
a serial of irreducible representations. In the following two sections,
we will list all the irreducible representations and discuss the corresponding
basis functions for both the singlet and triplet super-conducting
states in $LaOFeAs$.

\subsection{The irreducible representation and basis gap functions for spin singlet
super-conducting state in $LaOFeAs$.}

\begin{table*}[h]

\caption{Basis gap functions for the symmetry $D_{4h}$ (a). Spin-Singlet}

\label{Tab:basis_singlet} \begin{ruledtabular} \begin{tabular}{c|c|c|c|c|l|c}
 & ${\cal T}$  & P  & O  & S  & Basis  & Gap type \tabularnewline
\hline 
$A_{1g}$  & $+$  & $+$  & $+$  & $-$  & $\sigma_{0}(x^{2}+y^{2}),~\sigma_{1}(x^{2}+y^{2}),~\sigma_{0}(2z^{2}-x^{2}-y^{2}),~\sigma_{3}(xy),~\sigma_{1}(2z^{2}-x^{2}-y^{2})$  & I;II;III\tabularnewline
\hline 
$A_{2g}$  & $+$  & $+$  & $+$  & $-$  & $\sigma_{3}(x^{2}-y^{2})$  & II\tabularnewline
\hline 
$B_{1g}$  & $+$  & $+$  & $+$  & $-$  & $\sigma_{0}(x^{2}-y^{2}),~\sigma_{1}(x^{2}-y^{2})$  & II;III\tabularnewline
\hline 
$B_{2g}$  & $+$  & $+$  & $+$  & $-$  & $\sigma_{3}(x^{2}+y^{2}),~\sigma_{0}(xy),~\sigma_{3}(2z^{2}-x^{2}-y^{2}),~\sigma_{1}(xy)$  & II;III\tabularnewline
\hline 
$E_{g}$  & $+$  & $+$  & $+$  & $-$  & $\sigma_{0}(xz),~\sigma_{0}(yz),~\sigma_{3}(xz),~\sigma_{3}(yz),~\sigma_{1}(xz),~\sigma_{1}(yz)$  & II,III\tabularnewline
\hline 
$B_{1u}$  & $-$  & $-$  & $-$  & $-$  & $i\sigma_{2}(z)$  & II,III\tabularnewline
\hline 
$E_{u}$  & $-$  & $-$  & $-$  & $-$  & $i\sigma_{2}(y),~i\sigma_{2}(x)$  & II;III\tabularnewline
\end{tabular}\end{ruledtabular} 
\end{table*}

In the above table, we list all the basis gap functions for spin-singlet
super-conducting states in $LaOFeAs$, where $\sigma_{0}$ , $\sigma_{1}$,
$\sigma_{2}$ , $\sigma_{3}$ are $2\times2$ unit matrix and Pauli
matrices respectively, which form a complete basis for the orbital
space. On the first three columns we list the even-odd properties
of the basis function under the time reversal ($T$), spacial inversion
($P$), spin exchange ($S$) and orbital exchange ($O$), with the
plus and minus sign representing the even and odd functions under
the transformation respectively. The anti-symmetric condition for
the Fermion exchange requires $P\cdot S\cdot O=-1$, which is well
satisfied by each basis function. Due to the multi-band nature, the
system show some interesting feature which is different with the single
band case. Firstly because of the additional orbital degree of freedom,
it is possible to have odd parity basis functions for spin singlet
phase, as in representation $B_{1u}$ and $E_{u}$. Secondly in $A_{1g}$
and $B_{2g}$, the s-wave basis functions can mix with some d-wave
basis functions. This is because the special FS topology around the
$M$ point. The two electron type FS enclosing $M$ point form a 2-dimensional
representation for $D_{4h}$. If we only consider the pairing order
parameters within one particular band, the $D_{4h}$ symmetry will
be broken.

In the following, we will discuss in deal the super-conducting gap
functions for $A_{1g}$,which is the trivial representation of $D_{4h}$,
as an example. In this case, the super-conducting order parameter
can be written as the linear combination of the three in-plane basis
functions, which reads,

\[
\Delta_{s}^{11}(k)=a\left(cos(k_{x})+cos(k_{y})\right)+bsin(k_{x})sin(k_{y})\]

\[
\Delta_{s}^{22}(k)=a\left(cos(k_{x})+cos(k_{y})\right)-bsin(k_{x})sin(k_{y})\]

\[
\Delta_{s}^{12}(k)=\Delta_{s}^{21}(k)=c\left(cos(k_{x})+cos(k_{y})\right)\]

With the different value of $a$,$b$ and $c$, the super-conducting
gap may have following three different types.

\begin{enumerate}
\item Full energy gap, which is labeled by $\mathfrak{\mathrm{I}}$ in the
above table. When the intra-band s-wave component $a\neq0$, there
is finite energy gap right on the Fermi level for the whole FS. 
\item Energy gap with a nodal line along the $x$ and $y$ axises, which
is labeled by $II$ . In order to have nodal lines in this particular
representation, we must have $a=c=0$ and $b\neq0$. In this case
the super-conducting gap vanishes at four FS crossing points. 
\item Fermi arc which is labeled by $III$. With this type of gap structure,
there is a finite section of FS which remains ungaped, which is first
proposed by us in reference \cite{Dai} . This situation may happen
when only inter-band pairing strength $c\neq0$. 
\end{enumerate}
The second important representation for the singlet pairing states
in $LaOFeAs$ is $B_{1g}$ , within which the SC order parameters
are linear combinations of the intra-band and inter-band $d_{x^{2}-y^{2}}$states.
In this case the pairing is the strongest at four FS crossing points
and vanishes along the $x$ and $y$ axises. When only the inter-band
component is non-zero, the gap function will have a {}``Fermi arc''.

\subsection{The irreducible representation and basis gap functions for spin triplet
super-conducting state in $LaOFeAs$.}

\begin{table*}[h]

\caption{Basis gap functions for the symmetry $D_{4h}$ (b). Spin-Triplet}

\label{Tab:basis_triplet} \begin{ruledtabular} \begin{tabular}{c|c|c|c|c|l|c}
 & ${\cal T}$  & P  & O  & S  & Basis  & Gap type\tabularnewline
\hline 
$A_{1g}$  & $-$  & $+$  & $-$  & $+$  & $i\sigma_{2}[\vec{e}_{z}(x^{2}-y^{2})],~i\sigma_{2}[\vec{e}_{x}(yz)+\vec{e}_{y}(xz)]$  & III\tabularnewline
\hline 
$A_{2g}$  & $-$  & $+$  & $-$  & $+$  & $i\sigma_{2}[\vec{e}_{z}(xy)],~i\sigma_{2}[\vec{e}_{x}(xz)-\vec{e}_{y}(yz)]$  & III\tabularnewline
\hline 
$B_{1g}$  & $-$  & $+$  & $-$  & $+$  & $i\sigma_{2}[\vec{e}_{z}(x^{2}+y^{2})],~i\sigma_{2}[\vec{e}_{z}(2z^{2}-x^{2}-y^{2})],~i\sigma_{2}[\vec{e}_{x}(yz)-\vec{e}_{y}(xz)]$  & I;II;III\tabularnewline
\hline 
$B_{2g}$  & $-$  & $+$  & $-$  & $+$  & $i\sigma_{2}[\vec{e}_{x}(xz)+\vec{e}_{y}(yz)]$  & III\tabularnewline
\hline 
$E_{g}$  & $-$  & $+$  & $-$  & $+$  & $i\sigma_{2}[\vec{e}_{x}(x^{2}+y^{2})],~i\sigma_{2}[\vec{e}_{y}(x^{2}+y^{2})],~i\sigma_{2}[\vec{e}_{x}(2z^{2}-x^{2}-y^{2})],~i\sigma_{2}[\vec{e}_{y}(2z^{2}-x^{2}-y^{2})]$  & I;II;III\tabularnewline
 &  &  &  &  & $i\sigma_{2}[\vec{e}_{x}(x^{2}-y^{2})],~i\sigma_{2}[\vec{e}_{y}(x^{2}-y^{2})],~i\sigma_{2}[\vec{e}_{x}(xy)],~i\sigma_{2}[\vec{e}_{y}(xy)]$  & \tabularnewline
 &  &  &  &  & $i\sigma_{2}[\vec{e}_{z}(xz)],~i\sigma_{2}[\vec{e}_{z}(yz)]$  & \tabularnewline
\hline 
$A_{1u}$  & $+$  & $-$  & $+$  & $+$  & $\sigma_{0}[\vec{e}_{x}(x)+\vec{e}_{y}(y)],~\sigma_{0}[\vec{e}_{z}(z)],~\sigma_{3}[\vec{e}_{x}(y)+\vec{e}_{y}(x)],~\sigma_{1}[\vec{e}_{x}(x)+\vec{e}_{y}(y)],~\sigma_{1}[\vec{e}_{z}(z)]$  & I;II;III\tabularnewline
\hline 
$A_{2u}$  & $+$  & $-$  & $+$  & $+$  & $\sigma_{0}[\vec{e}_{x}(y)-\vec{e}_{y}(x)],~\sigma_{3}[\vec{e}_{x}(x)-\vec{e}_{y}(y)],~\sigma_{1}[\vec{e}_{x}(y)-\vec{e}_{y}(x)]$  & I;II;III\tabularnewline
\hline 
$B_{1u}$  & $+$  & $-$  & $+$  & $+$  & $\sigma_{0}[\vec{e}_{x}(x)-\vec{e}_{y}(y)],~\sigma_{3}[\vec{e}_{x}(y)-\vec{e}_{y}(x)],~\sigma_{1}[\vec{e}_{x}(x)-\vec{e}_{y}(y)]$  & I;II;III\tabularnewline
\hline 
$B_{2u}$  & $+$  & $-$  & $+$  & $+$  & $\sigma_{0}[\vec{e}_{x}(y)+\vec{e}_{y}(x)],~\sigma_{3}[\vec{e}_{x}(x)+\vec{e}_{y}(y)],~\sigma_{3}[\vec{e}_{z}(z)],~\sigma_{1}[\vec{e}_{x}(y)+\vec{e}_{y}(x)]$  & I;II;III\tabularnewline
\hline 
$E_{u}$  & $+$  & $-$  & $+$  & $+$  & $\sigma_{0}[\vec{e}_{x}(z)],~\sigma_{0}[\vec{e}_{y}(z)],~\sigma_{0}[\vec{e}_{z}(x)],~\sigma_{0}[\vec{e}_{z}(y)],~\sigma_{3}[\vec{e}_{x}(z)],~\sigma_{3}[\vec{e}_{y}(z)],~$  & \tabularnewline
 &  &  &  &  & $\sigma_{3}[\vec{e}_{z}(x)],~\sigma_{3}[\vec{e}_{z}(y)],~\sigma_{1}[\vec{e}_{x}(z)],~\sigma_{1}[\vec{e}_{y}(z)],~\sigma_{1}[\vec{e}_{z}(x)],~\sigma_{1}[\vec{e}_{z}(y)]$  & I;II;III\tabularnewline
\end{tabular}\end{ruledtabular} 
\end{table*}

In the above table, we list all the basis gap functions for spin-triplet
super-conducting states in $LaOFeAs$. Since the total spin of the
Cooper pairs generate extra degree of freedom, the situation for the
spin triplet SC states is more complicated. In the present study,
we will only focus on the possible gap structure. The first five representations
with even parity are purely inter-band pairing states. In the previous
paper, we have discussed the gap structure of inter-band s-wave and
$d_{x^{2}-y^{2}}$-wave states , which are listed in the$B_{1g}$
and $A_{1g}$ representations respectively. For the inter-band s-wave
state with order parameter $i\sigma_{2}[\vec{e}_{z}(x^{2}+y^{2})]$,
the system will have full gap at the FS when the amplitude of the
pairing order parameter is much bigger than the maximum splitting
of the two bands at the FS and will have {}``Fermi Arc'' behavior
otherwise\citep{Dai}. And as we have discussed in the previous paper,
for the inter-band $d_{x^{2}-y^{2}}$ state, the {}``Fermi Arc''
will always appear around the diagonal line. The basis functions in
the$B_{2g}$ and $A_{2g}$ representations describe the inter-band
pairing states along the c-axis, which are very unlikely for a layered
compound like $LaOFeAs$.

The basis functions in the rest of the triplet representations mainly
describe the intra-band p-wave pairing states. But they can be mixed
with the corresponding inter-band p-wave pairing states. The $A_{1u}$
representation contains three in-plane pairing states, namely $\sigma_{0}[\vec{e}_{x}(x)+\vec{e}_{y}(y)]$,
$\sigma_{3}[\vec{e}_{x}(y)+\vec{e}_{y}(x)]$ and $\sigma_{1}[\vec{e}_{x}(x)+\vec{e}_{y}(y)]$.
The first one is quite well known in the Helium III system as the
BW phase. While in the present two-band system,very interestingly
the BW phase here may be mixed with two other states. The first one
is an intra-band pairing state with the order parameter $\overrightarrow{d}=\sigma_{3}[\vec{e}_{x}(y)+\vec{e}_{y}(x)]$,
and another one is an inter-band pairing state with $\overrightarrow{d}=\sigma_{1}[\vec{e}_{x}(x)+\vec{e}_{y}(y)]$,
which is nothing but the inter-band BW phase. Although the intra-band
BW phase itself has a full energy gap on the FS, it may acquire line
nodes or {}``Fermi Arc'' after mixing with the intra-band BW phase.
We will discuss the physical properties of this state in detail in
a separated paper. The behavior of the energy gaps for the $A_{2u}$,$B_{1u}$
and $B_{2u}$ states are quite similar with that of $A_{1u}$ states.
Another important representation for the triplet states is $E_{u}$,
which contains the ABM phase in Helium III system. The intra-band
ABM phase, which is proposed by M. Sigrist and M. Rice for the super-conductivity
in $SrRuO_{4}$, can be obtained by linear combination of two basis,
namely $\sigma_{0}[\vec{e}_{z}(x)]$ and $\sigma_{0}[\vec{e}_{z}(y)]$.
The intra-band ABM phase breaks the time reversal symmetry and also
can mix with the inter-band ABM phase. Similarly for purely inter-band
ABM phase, the {}``Fermi Arc'' behavior will appear.

\section{conclusions}

In the present study, we classified the possible super-conducting
order parameters for $LaOFeAs$ systems respect to all the crystal
symmetry including time reversal, inversion and point group symmetry.
Since the spin-orbital coupling in the system is weak, we can still
divide the possible super-conducting order parameters into spin singlet
and triplet two classes. The for each class we give the detailed classification
table with each class forming an irreducible representation. We also
discussed the possible gap structure for each class of super-conducting
phases. In the $LaOFeAs$ system, partly due to its multi-band nature,
there are three possible types of gap functions, namely full gap,
nodal gap and {}``Fermi Arc'' type. The different types of super-conducting
gap will have very different thermal dynamic and super-conducting
properties in low temperature, i.e. the specific heat and superfluid
density. The super-conducting state with full gap, like s-wave, will
have exponential temperature dependence in low temperature for specific
heat and superfluid density. If there are line nodes on the FS, both
the specific heat and superfluid density will show power law temperature
dependence. While for super-conducting state with finite {}``Fermi
Arc'', the low temperature behavior of specific heat appears like
there are some {}``residue density of states'' on the Fermi level.
Currently there are some experimental evidence indicating the existence
of low energy quasi-particles in the super-conducting phase\citep{Wen,Wang,NMR2}.
But we still need further experimental data to make the conclusion. 

We acknowledge valuable discussions with Y. P. Wang, N. L. Wang, J.R.
Shi, F.C. Zhang, Lu Yu and Z.B. Su. This work is supported by NSF
of China and 973 project of the Ministry of Science and Technology
of China..


\begin{references}

\bibitem{LOFS} Y. Kamihara, et. al., J. Am. Chem. Soc., (doi:10.1021/ja800073m),
(2008).

\bibitem{COFS} G. F. Chen, Z. Li, D. Wu, G. Li, W. Z. Hu, J. Dong,
P. Zheng, J. L. Luo, and N. L. Wang, cond-mat/0803.3790v1, (2008).

\bibitem{SOFS} X. H. Chen, T. Wu, G. Wu, R. H. Liu, H. Chen and D.
F. Fang, cond-mat/0803.3603v1, (2008).

\bibitem{POFS} Z. A. Ren, J. Yang, W. Lu, W. Yi, G. C. Che, X. L.
Dong, L. L. Sun, Z. X. Zhao, cond-mat/0803.4283v1, (2008).

\bibitem{SOFS2} Zhi-An Ren, Wei Lu, Jie Yang, Wei Yi, Xiao-Li Shen,
Zheng-Cai Li, Guang-Can Che, Xiao-Li Dong, Li-Ling Sun, Fang Zhouand
Zhong-Xian Zhao, cond-mat/0804.2053, (2008).

\bibitem{GOFS} Cao Wang, Linjun Li, Shun Chi, Zengwei Zhu, Zhi Ren,
Yuke Li, Yuetao Wang, Xiao Lin, Yongkang Luo, Xiangfan Xu, Guanghan
Cao, and Zhu'an Xu, cond-mat/0804.4290, (2008).

\bibitem{SOxFS2}Zhi-An Ren, Guang-Can Che, Xiao-Li Dong, Jie Yang,
Wei Lu, Wei Yi, Xiao-Li Shen, Zheng-Cai Li, Li-Ling Sun, Fang Zhou,
and Zhong-Xian Zhao, cond-mat/0804.2582, (2008).

\bibitem{Hole} H. H. Wen, and et.al., cond-mat/0803.3021v1 (2008).

\bibitem{SDW} J. Dong, H. J. Zhang, G. Xu, Z. Li, G. Li, W. Z. Hu,
D. Wu, G. F. Chen, X. Dai, J. L. Luo, Z. Fang, and N. L. Wang, cond-mat/0803.3426v1,
(2008).

\bibitem{Neutron} Clarina de la Cruz, Q. Huang, J. W. Lynn, Jiying
Li, W. Ratcliff II, J. L. Zarestky, H. A. Mook, G. F. Chen, J. L.
Luo, N. L. Wang,and Pengcheng Dai, cond-mat/0804.0795, (2008).

\bibitem{Neutron2} M. A. McGuire, A. D. Christianson, A. S. Sefat,
R. Jin, E. A. Payzant, B. C. Sales, M. D. Lumsden and D. Mandrus,cond-mat/0804.0796,
(2008).

\bibitem{SDW-Cal} Fengjie Ma, Zhong-Yi Lu and Tao Xiang, cond-mat/0804.3370,
(2008).

\bibitem{SDW-Cal2} T. Yildirim, cond-mat/0804.2252, (2008).

\bibitem{NMR} K. Ahilan, F. L. Ning, T. Imai, A. S. Sefat, R. Jin,
M.A. McGuire, B.C. Sales, D. Mandrus, cond-mat/0804.4026, (2008).

\bibitem{NMR2} Yusuke Nakai, Kenji Ishida, Yoichi Kamihara, Masahiro
Hirano, Hideo Hosono, cond-mat/0804.4765, (2008).

\bibitem{Xu} G. Xu, W. Ming, Y. Yao, X. Dai, S. C. Zhang, and Z.
Fang, cond-mat/0803.1282v2, (2008).

\bibitem{Singh} D. J. Singh and M. H. Du, cond-mat/0803.0429v1, (2008).

\bibitem{Mazin} I.I. Mazin, D.J. Singh, M.D. Johannes, and M.H. Du,
cond-mat/0803.2740.v1, (2008).

\bibitem{Kuroki} K. Kuroki, and et.al., cond-mat/0803.3325v1 (2008).

\bibitem{LOFP-cal} S. Lebégue, Phys. Rev. B 75, 035110 (2007).

\bibitem{Kotliar} K. Haule, J. H. Shim, and G. Kotliar, cond-mat/0803.1279v1
(2008).

\bibitem{HJZhang} H. J. Zhang, et. al. cond-mat/0803.4487.

\bibitem{Dai} X. Dai, Z. Fang, Y. Zhou. F. C. Zhang, cond-mat/0809.3982v1
(2008).

\bibitem{Shoucheng} Xiao-Liang Qi, S. Raghu, Chao-Xing Liu, D. J.
Scalapino, Shou-Cheng Zhang, cond-mat/0804.4332 (2008).

\bibitem{Xiaogang}Patrick A. Lee, Xiao-Gang Wen, cond-mat/0804.1739
(2008).

\bibitem{Zidan}Zi-Jian Yao, Jian-Xin Li, Z. D. Wang, cond-mat/0804.4166
(2008).

\bibitem{Qimiao}Qimiao Si, Elihu Abrahams , cond-mat/0804.2480 (2008).

\bibitem{Scalapino}T.A. Maier, D.J. Scalapino, cond-mat/0805.0316
(2008).

\bibitem{Sigrist}M. Sigrist and K. Ueda, Rev. Mod. Phys. 63, 239
(1991).

\bibitem{Leggett}A. J. Leggett, Rev. Mod. Phys., 47, 331 (1975).

\bibitem{Wang} H. Yang, et al., cond-mat/0803.0623; G. F. Chen, et
al., cond-mat/0803.0128 (2008).

\bibitem{Wen} L. Shan, et al., cond-mat/0803.2405v2 (2008); X. Zhu,
et al., cond-mat/0803.1288 (2008); G. Mu, et al., cond-mat/ 0803.0928
(2008).

\end{references}
\end{document}